\documentclass[12pt,a4paper]{article}
\usepackage[utf8]{inputenc}
\usepackage[english]{babel}
\usepackage{indentfirst}
\usepackage{misccorr}
\usepackage{graphicx}
\usepackage{amsmath}
\usepackage{amssymb}
\usepackage{amsfonts}
\usepackage{amsthm}
\usepackage{tocloft}
\usepackage{tikz}
\usepackage{pgfplots}
\usetikzlibrary{calc}
\usetikzlibrary{intersections}
\usepackage{xcolor}
\usepackage[left=20mm, top=20mm, right=15mm, bottom=20mm]{geometry}

\theoremstyle{plane}
\numberwithin{equation}{section}

\theoremstyle{definition}

\newtheorem*{definition*}{Определение}

\title{Transformation of synchronization picture with\\ a change of an external pulse action direction in\\ three-dimensional R\"ossler system}
\author{P.Yu.~Panteleeva, N.V.~Stankevich\\
\textit{HSE University}}
\date{}

\begin{document}

\maketitle
\small

{\bf Abstract.} We consider the non-autonomous three-dimensional R\"ossler system under the external pulse action. In this work we describe changing of synchronization picture depending on the direction of the external action. Complex oscillatory modes, including quasi-periodic, chaotic and hyperchaotic, initiated by an external force are studied.



\section{Introduction}

Processes that are described by systems under the action of external forces occur in many areas of physics, biology, chemistry and other natural sciences \cite{Christiansen, Safonov, Tij, Zi-zhen, Asokan, G-Odelin, Pipin}. The paper \cite{Christiansen} investigates the action of random external influences on the occurrence and dynamic characteristics of a transonic flutter of a wing with a certain configuration. The work \cite{Tij} considers the dynamics of a gas flow in a layer under the action of a constant external force. In \cite{Zi-zhen} the dynamic mechanisms of crop growth under the influence of various external factors (light, temperature, soil, nutrients, etc.) are considered. The reactions of the growth rate to changes in these external influences have been studied. The paper \cite{Asokan} investigates the properties of a dilute suspension of Brownian dipole spheroids in a simple shear flow under the action of constant and periodic external forces. The paper \cite{G-Odelin} presents the construction of a new class of solutions to the non-equilibrium Boltzmann equation under the action of an external force.

An important class of tasks are related to problems in which the behavior of systems under the influence of periodic pulses of constant amplitude is studied \cite{Gonzalez, Ullmann, KT, Obodan}. In the case when an autonomous system demonstrates periodic self-oscillations, then a classical synchronization phenomenon is observed, when the natural frequency of the self-oscillating system adjusts to external action. In the case when an autonomous system demonstrates different dynamics, for example, chaotic, then we can talk about stabilization (or control) of chaotic behavior with the help of external action \cite{KST2006, KST20062, KST2007, Loskutov}. In this case, the direction of the external action also plays an important role; from the point of view of physical meaning, this can be the influence of various factors of the dynamical system.

The Andronov-Hopf bifurcation is the simplest bifurcation of the limet cycle birth in a flow dynamical system, which can occur in a system with a minimum dimension of the phase space equal to two \cite{Andronov}. In this case, the birth of a limit cycle is associated with the transformation of the equilibrium state from a stable focus to an unstable focus. With an increase in the phase space dimension to three, as a result of the Andronov-Hopf bifurcation, a saddle focus or an unstable focus can be born from a stable focus and and a saddle focus, respectively. At the same time, for a system with a three-dimensional phase space, the bifurcation essentially occurs on a plane, the saddle-focus after the Andronov-Hopf bifurcation has a two-dimensional unstable manifold and a one-dimensional stable manifold. When an external force is added to such a system, the direction of the force can significantly change the synchronization picture, depending on whether the vector of the external force lies inside the plane of the two-dimensional manifold or not. It was shown earlier \cite{KST2006, KST20062}, that the synchronization picture for a saddle-focus system really depends on the direction of the external pulse action. At the same time, cases were shown when the influence was directed only in the direction of the axes of the external influence, a direction was identified that significantly changes the picture. A detailed analysis of the transformation of the synchronization picture was not carried out, and the mechanisms of destruction of the picture were not described. The purpose of this work is to develop a mathematical model that allows us to control the direction of external influence and to study the transformation of the synchronization picture with a smooth change of the external force direction.

The work is structured as follows. In section 2 we present the objects and methods of study: an autonomous system, its parameter space, we will select working parameters; a non-autonomous system that allows us to change the direction of action of periodic impulses. In Section 3 we discuss the several cases of the direction of action and describes the synchronization picture for each case. Section 4 presents illustrations of the transformation of the picture when changing the parameters responsible for the external signal direction.


\section{Object of study}

As an object of study, we choose the R\"ossler system, the dynamics of which develops on the basis of saddle-focus equilibrium states. It is described by a system of three ordinary differential equations of the form
\begin{equation}
    \begin{array}{l l l}
    \dot x = -y-z\\
    \dot y = x+py\\
    \dot z = q+z(x-r)
    \end{array}
    \label{roessler}
\end{equation}
where $x,y,z$ are dynamical variables, $p,q,r\in\mathbb{R}$ are parameters that determine the dynamics of the system. It is easy to see that the R\"ossler system (\ref{roessler}) has two or one equilibrium states:
$$
    (x^0_\pm,y^0_\pm,z^0_\pm)=\left(\frac{r\pm\sqrt{r^2-4pq}}{2},-\frac{r\pm\sqrt{r^2-4pq}}{2p},\frac{r\pm\sqrt{r^2-4pq}}{2p}\right)
$$
\begin{figure}[!h]
\center{\includegraphics[width=0.90\linewidth]{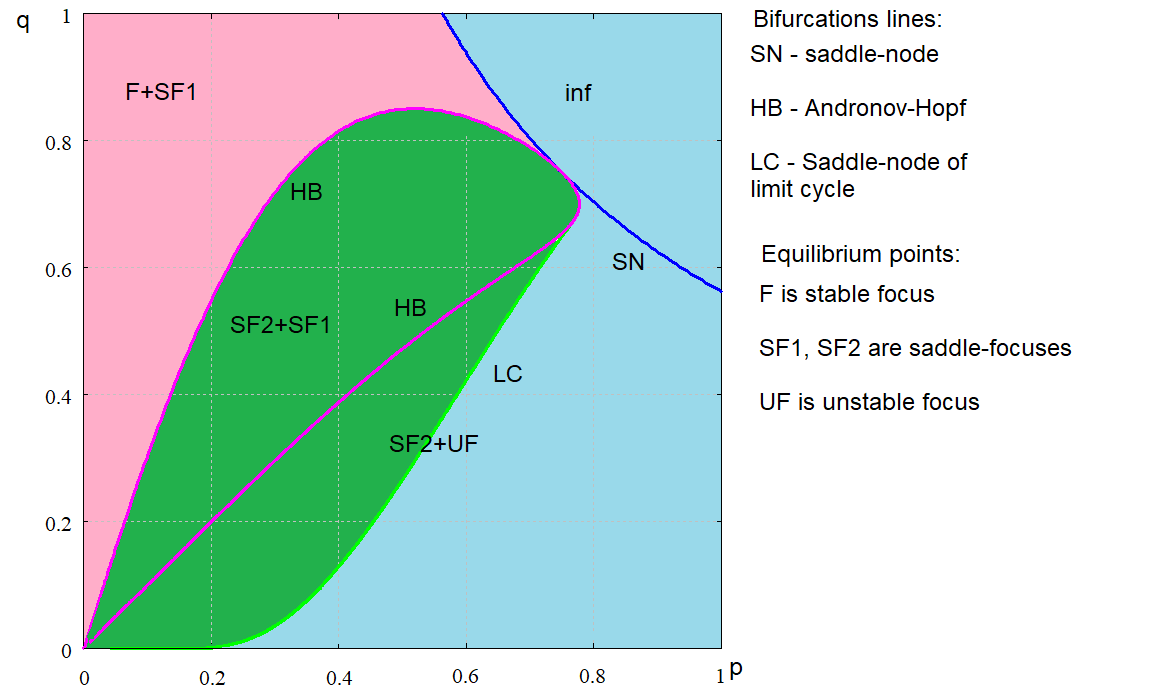}}
\caption{The parameter plane, on which various types of equilibrium states and dynamic regimes of the R\"ossler system are classified at $r = 1.5$. Pink color is stable fixed point; blue color is divergence; green is stable limit cycle}
\label{fig1:chartEP}
\end{figure}
When $r^2-4pq>0$ there are two fixed points in the phase space of system (\ref{roessler}), at the moment $r^2-4pq=0$ a saddle-node bifurcation occurs, as a result of which the fixed points merge into one semi-stable one and then, for $r^2-4pq<0$, disappear. For $r^2-4pq>0$, for fixed $p$ and $q$ and increasing parameter $r$, the equilibrium states experience Andronov-Hopf bifurcations. Figure~\ref{fig1:chartEP} shows a bifurcation diagram obtained in the XPPAUT \cite{ermentrout} numerical bifurcation analysis package. Three areas can be distinguished on the map, where there is a stable equilibrium, a stable limit cycle, and also an area where the trajectories run away to infinity.In this work we will fix the parameters such that a stable limit cycle will be implemented in the autonomous system: $p=0.1$, $q=0.2$, $r=1.5$.

Let us set the external action as a periodic sequence of $\delta$-pulses. The influence of such pulses directed along the $x$ axis and along the $z$ axis was previously considered in \cite{KST2006, KST20062}. In this case, the external action was specified using an additional term of a special form:
$$
F(t)=A\sum\limits_{n=-\infty}^{+\infty}\delta(t-nT),
$$
where $A$ and $T$ are amplitude and period of the external pulse action, $\delta(s)$ is a single impulse at the point $s$. In the works described above, it is shown that the picture of synchronization of the non-autonomous R\"ossler system in these two cases differs significantly. However, the features of the transformation of the synchronization pattern with a smooth change in the external action have not been studied.

\begin{figure}[!h]
\begin{minipage}[h]{0.32\linewidth}
\center{\includegraphics[width=0.90\linewidth]{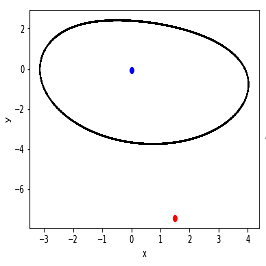}}
\end{minipage}
\begin{minipage}[h]{0.32\linewidth}
\center{\includegraphics[width=0.90\linewidth]{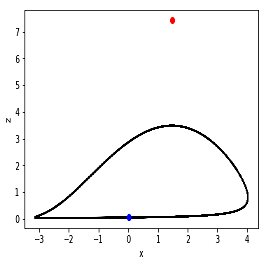}}
\end{minipage}
\begin{minipage}[h]{0.32\linewidth}
\center{\includegraphics[width=0.90\linewidth]{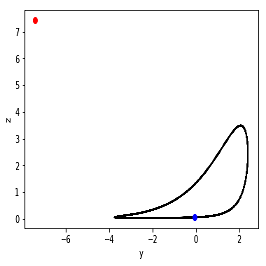}}
\end{minipage}
\caption{Two-dimensional projections of the attractor of the autonomous R\"ossler system for $p = 0.1$, $q = 0.2$, $r = 1.5$. The blue and red dots correspond to the equilibrium positions $P_1$ and $P_2$, respectively.}
\label{phase portraits}
\end{figure}

In order to be able to track the features of the transformation of the synchronization pattern when the direction of the external action changes, we will consider a model that allows us to trace the change in the dynamics of the system under the influence of pulses directed not only along all coordinate axes, but also in intermediate directions. To do this, we will introduce the angles $\varphi$ and $\psi$, with the help of which we will control the direction of the external action. The modified system will have the following form:	
\begin{equation}
    \begin{cases}
    \dot x = -y-z+F(t)\sin\varphi\cos\psi\\
    \dot y = x+py+F(t)\sin\varphi\sin\psi\\
    \dot z = q+z(x-r)+F(t)\cos\varphi,
    \end{cases} 
    \label{angles}
\end{equation}
where $F(t)$ is a pulse action with fixed amplituse and period, $\varphi$ and $\psi$ are the angles that determine the direction of the external action in the phase space. For the values of the parameters fixed by us, the autonomous system has a stable limit cycle, most of which lies in the $(x,y)$ plane and only a relatively small part juts out in the direction of the $z$ axis. Two-dimensional projections of this cycle are shown in Fig.\ref{phase portraits}.

In this case, there are two fixed points in the phase space of the system: $P_1$ is a saddle-focus with a two-dimensional unstable manifold with coordinates $(0.0135, -0.0673, 0.0673)$, characterized by the following eigenvalues: $\lambda_{1,2}=0.0837 \pm i1.0031$, $\lambda_3=-1.3549$; $P_2$ is an unstable focus with coordinates $(1.4865, -7.4327, 7.4327)$, characterized by the following eigenvalues: $\lambda_{1,2}=0.0059\pm i2.9031$, $\lambda_3=0.1748$.

On fig. \ref{phase portraits} it is clearly seen that the projection of the attractor onto the $(x, y)$ plane is close to a circle, which means that applying pulses in the directions along the $x$ axis and the $y$ axis should give approximately the same result , close to the classical synchronization pattern (circle mapping) \cite{ding}, in this case, the deviation of the external signal in the direction of the $z$ axis should significantly change the pattern of regimes. We also note that the saddle-focus equilibrium is inside the limit cycle, while the unstable focus is far from the attractor. By changing the angles $\varphi$ and $\psi$ from zero to $\pi$, we will change the direction of the external force on the system, which will allow us to trace how the dynamics of the system changes in this case.

\section{Limit cases}

First, consider the limiting cases when the action of an external force is directed along the coordinate axes: a) – the action is directed along the $x$ axis: $\varphi = \pi/2$, $\psi=0$; b) – the impact is directed along the $y$ axis: $\varphi = \pi/2$, $\psi = \pi/2$; c) – the action is directed along the $z$ axis: $\varphi = 0$, the angle $\psi$ is arbitrary. These cases were previously discussed in \cite{KST20062}. However, the method of dynamical regime maps was used as the main research tool, which does not allow one to distinguish between quasi-periodic and chaotic oscillations. At the same time, it was shown in \cite{ND09}, that these systems can demonstrate not only quasi-periodic and chaotic oscillations, but also torus bifurcations. In this connection, in this work, as a tool, we will also use the Lyapunov indicator map method. On Fig. \ref{xyz} and Fig. \ref{Lyap_xy} maps of dynamical regimes and maps of Lyapunov exponents for three limiting cases of the direction of external influence are presented.

The construction of maps of dynamical regimes was carried out as follows. The parameter plane was scanned with a small step, for each point the attractors in the Poincar\'e stroboscopic section were analyzed: the number of fixed points in the section was counted, the period of the attractor was determined in accordance with this number, and the point on the plane was painted in one color or another (the palette in Fig. \ref{xyz}). With more than 10 points, the attractor was considered irregular (black color on the map). Gray color on the maps marks the areas of trajectory run away to infinity (this mode was determined when the dynamic variables reached the absolute value of 10000). The maps of Lyapunov exponents were calculated in a similar way, but only for each point the full range of Lyapunov exponents was calculated using the Benettin algorithm and Gram-Schmidt orthogonalization \cite{Benettin}. The R\"ossler system (\ref{angles}) is non-autonomous, characterized by four Lyapunov exponents, one of which is always equal to zero. Depending on the signature of the three remaining indicators, we classified the following types of dynamic behavior:
\begin{itemize}
   \item periodic oscillations ($0>\Lambda_1 > \Lambda_2 > \Lambda_3$);
     \item two-frequency quasi-periodic oscillations ($\Lambda_1 = 0, 0 > \Lambda_2 > \Lambda_3$);
     \item three-frequency quasi-periodic oscillations ($\Lambda_1 = \Lambda_2 = 0, 0 > \Lambda_3$);
    \item chaotic oscillations ($\Lambda_1 > 0, 0 > \Lambda_2 > \Lambda_3$);
      \item hyperchaotic oscillations ($\Lambda_1 > \Lambda_2 > 0, 0 > \Lambda_3$);
      \item chaotic oscillations with additional zero Lyapunov exponent ($\Lambda_1 >, \Lambda_2 = 0, 0 > \Lambda_3$).
\end{itemize}

\begin{figure}[!h]
\begin{minipage}[h]{0.33\linewidth}
\center{\includegraphics[width=0.98\linewidth]{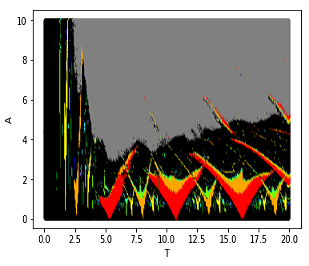}}
\end{minipage}
\begin{minipage}[h]{0.33\linewidth}
\center{\includegraphics[width=0.98\linewidth]{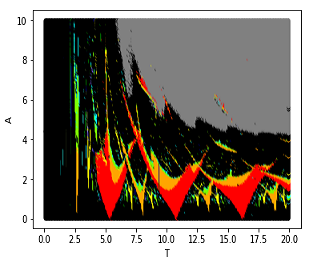}}
\end{minipage}
\begin{minipage}[h]{0.33\linewidth}
\center{\includegraphics[width=0.95\linewidth]{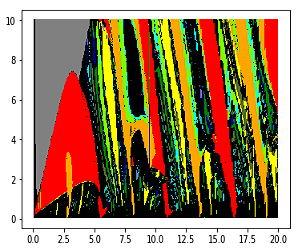}}
\end{minipage}
\end{figure}
\begin{figure}[!h]
\center{\includegraphics[width=0.90\linewidth]{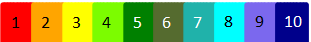}}
\caption{Charts of dynamical regimes of non-autonomous R\"ossler system (\ref{angles}) with $p = 0.1$, $q = 0.2$, $r = 1.5$ and different directions of external action: a) $\varphi = \pi/2$, $\psi=0$ is action along the axis $x$; b) $\varphi = \pi/2$, $\psi = \pi/2$ is action along the axis $y$; c) $\varphi = 0$ is action along the axis $z$.}
\label{xyz}
\end{figure}

\begin{figure}[!h]
\begin{minipage}[h]{0.33\linewidth}
\center{\includegraphics[width=0.98\linewidth]{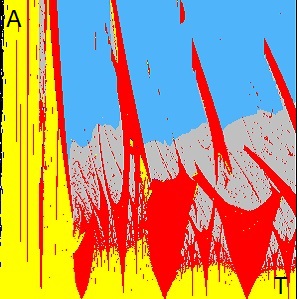}}
\end{minipage}
\begin{minipage}[h]{0.33\linewidth}
\center{\includegraphics[width=0.98\linewidth]{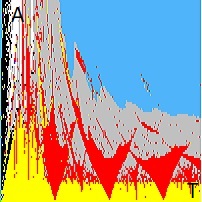}}
\end{minipage}
\begin{minipage}[h]{0.33\linewidth}
\center{\includegraphics[width=0.98\linewidth]{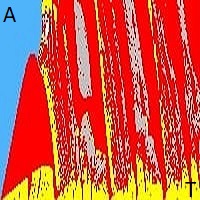}}
\end{minipage}
\caption{Charts of Lyapunov exponents for the systems with pulse action along the axis $x$ (left), $y$ axes (center), and $z$ axes (right). The abscissa shows the periods $T$, the ordinate shows the amplitudes $A$.}
\label{Lyap_xy}
\end{figure}

When an external influence is added to the system, if the autonomous system exhibits periodic self-oscillations, we can talk about the classical phenomenon of synchronization. At the same time, on the plane of parameters of the external signal (period, $T$ - amplitude, $A$), a characteristic pattern of synchronization is observed: from the points of a rational ratio of the frequency of natural oscillations of the system and external influence at zero amplitude, synchronization tongues emerge.Quasi-periodic fluctuations are observed between tongues. As the signal amplitude grows, the picture becomes more complicated; various scenarios for the development of chaotic dynamics and overlapping of tongues can be realized. It is clearly seen in Fig.~\ref{xyz} that the patterns that appear when exposed to the $x$ and $y$ axes are largely classical.Synchronization tongues are well traced. As the amplitude increases, cascades of period doubling bifurcations occur within tongues with the development of chaos. The overlapping line of synchronization tongues is also well traced, which corresponds to the loss of smoothness of the torus and the development of chaos in accordance with the Afraimovich-Shilnikov scenario. At the same time, on the map of Lyapunov exponents, we see mostly chaotic fluctuations with one positive Lyapunov exponent. An interesting device has an area of small periods of external influence (high-frequency influence). Here we see that for large amplitudes the quasi-periodic dynamics is preserved for the case of action along the $x$ axis. For the impact along the $y$ axis, we see the development of various types of chaos, including hyperchaos.

Significant differences in the synchronization pattern arise in the case of action along the $z$ axis. At small amplitudes, we can detect synchronization tongues, however, as the amplitude increases inside the tongues, it is impossible to track the cascades of period doubling bifurcations. Tongues alternate with areas of quasi-periodic fluctuations. The regions of high-frequency oscillations are fundamentally different. At small periods of external action, quasi-periodic oscillations are observed, which are limited by the lines of the reverse Neimark-Sacker bifurcation, as a result of which periodic self-oscillations appear. The region of periodic self-oscillations has a clear threshold in terms of the amplitude of the external action, when it is exceeded, the trajectories run away to infinity. As the external action period tends to zero, this threshold also tends to zero, which is the opposite of the synchronization pattern for the other two limiting cases.

In the next section, we will consider the gradual transformation of the synchronization pattern when the direction of the external force is changed by changing the angles $\varphi$ and $\psi$ in the system (\ref{angles}).

\section{Transformation of the picture when changing the direction of external action}

Let us now consider the cases of impact on the system of pulses applied in intermediate directions. To do this, we will fix one of the corners and sequentially change the other, tracing how the map of dynamic modes is transformed in this case. As a result of such a study, the following observations can be made: (1) in the absence of a component directed along the $z$ axis in the signal, the picture, as expected, is transformed insignificantly; (2) with a significant contribution of the $z$-component, the picture is close to that which arises in Fig. 2c, however, as the angle $\varphi$ increases from zero to $\pi/2$, the region of irregular motions disappears at small periods.
\begin{figure}[!h]
\begin{minipage}[h]{0.48\linewidth}
\center{\includegraphics[width=0.98\linewidth]{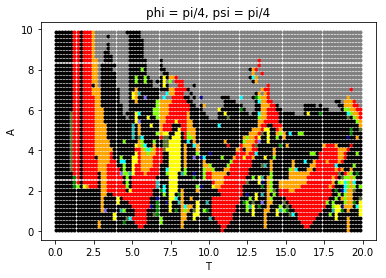}}
\end{minipage}
\begin{minipage}[h]{0.48\linewidth}
\center{\includegraphics[width=0.98\linewidth]{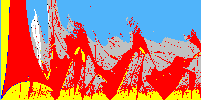}}
\end{minipage}
\caption{The chart of dynamical regimes and the chart of Lyapunov exponents corresponding to the value of the angles $\varphi=\psi=\pi/4$ in the \ref{angles} system}
\label{pi_4}
\end{figure}

As an example, consider separately the case $\varphi=\psi=\pi/4$ (Fig. \ref{pi_4}). The figure shows that for such values of the angles there is no chaotic region for small periods of action. In addition, the area of hyperchaos is visible on the chart of Lyapunov exponents.

   
\section{Conclusion}
Systems with saddle-focus equilibrium states are widely used in various applied problems, including biophysical models: neurons, models of protein interactions, etc. \cite{Shilnikov,Stanoev}. In the framework of this work, we study the features of the pattern of regimes in a system with a saddle-focus when changing the direction of the external signal. The direction of the external pulse action significantly affects the synchronization pattern. When a signal is applied in the plane where the two-dimensional unstable manifold is located (the unstable focus is untwisted), a picture is observed that is close to the classical picture of synchronization. The peculiarity of the saddle-focus character is manifested for small periods of external action. An increase in the region of quasi-periodic oscillations is observed, with chaos appearing chaos with a different signature of the spectrum of Lyapunov exponents. A change in direction from the specified plane leads to the destruction of the classical picture. Secondary Neimark-Sacker bifurcations and two-frequency tori of a higher period appear. Complex dynamics in the high-frequency area are completely destroyed 

 \section{Acknowledgments}
The work is supported by the grant of the Russian Science Foundation (project No. 20-71-10048). This research was supported in part through computational resources of HPC facilities at HSE University


\begin{thebibliography}{}
\bibitem{Andronov}
Andronov A. A., Vitt A. A., Khaikin S. E. Theory of Oscillators: Adiwes International Series in Physics. – Elsevier, 2013. – v. 4.
\bibitem{Asokan}
Asokan K., Ramamohan T. R. The rheology of a dilute suspension of Brownian dipolar spheroids in a simple shear flow under the action of an external force //Physics of Fluids. – 2004. – v. 16(2). – p. 433-444.
\bibitem{Benettin}
Giancarlo Benettin, Luigi Galgani, Antonio Giorgilli, Jean-Marie Strelcyn, Lyapunov Characteristic Exponents for smooth dynamical systems and for hamiltonian systems; a method for computing all of them. Part 1: Theory // Meccanica, v. 15, pp. 9–20 (1980)
\bibitem{Christiansen}
Christiansen L. E. et al. Nonlinear characteristics of randomly excited transonic flutter //Mathematics and computers in simulation. – 2002. – v. 58(4-6). – p. 385-405.
\bibitem{ding}
Ding E. J., Hemmer P. C. Winding numbers for the supercritical sine circle map // Physica D: Nonlinear Phenomena. – 1988. – v. 32(1) – p. 153-160
\bibitem{ermentrout}
Ermentrout B. Simulating, Analyzing and Animating Dynamical Systems: A guide to XPPAUT for researcher and students (SIAM, Philadelphia, 2002)
\bibitem{Gonzalez}
Gonzalez D. L., Piro O. Chaos in a nonlinear driven oscillator with exact solution //Physical Review Letters. – 1983. – v. 50(12). – p. 870.
\bibitem{G-Odelin}
Gu\'ery-Odelin D. et al. Nonequilibrium solutions of the boltzmann equation under the action of an external force //Physical review letters. – 2014. – p. 112(18). – p. 180602.
\bibitem{Kuz}
Kuznetsov A. P. Dynamical Chaos. – 2001. (in Russian)
\bibitem{KST2006}
Kuznetsov A. P., Stankevich N. V., Tyuryukina L. V. Features of the pattern of synchronization by pulses in a system with a three-dimensional phase space on the example of the R\"ossler system. Applied nonlinear dynamics. – 2006. – v. 14(6), – p. 43-53. (in Russian)
\bibitem{KST20062}
Kuznetsov A. P., Stankevich N. V., Tyuryukina L. V. Features of pulsed synchronization of an autooscillatory system with a three-dimensional phase space //Technical physics letters. – 2006. – v. 32(4). – p. 343-346.
\bibitem{KST2007}
Kuznetsov A.P., Stankevich N.V., Turukina L.V. Picture of Pulsed Synchronization in the Dmitriev – Kislov Generator//Nonlinear Phenomena in Complex Systems. – 2007. – v. 10(4). – p. 407-412.
\bibitem{ND09}
Kuznetsov A.P., Stankevich N.V., Turukina L.V., Stabilization by External Pulses and Synchronous Response in the R\"ossler System up to the Saddle-Node Bifurcation Threshold. Nonlinear Dynamics, 5(2), 2009, p. 253-264 (in Russian)
\bibitem{KT}
Kuznetsov A.P., Turukina L.V. Synchronization of the Van der Pol-Duffing self-oscillating system with short pulses // Izv. universities. Applied nonlinear dynamics. – 2004. – v. 12(5). – p. 16-31. (in Russian)
\bibitem{Loskutov}
Loskutov A. Y., Cheremin R. V., Vysotskii S. A. Stabilization of turbulent dynamics in excitable media by an external point action //Doklady Physics. – Nauka/Interperiodica, 2005. – v. 50(10). – p. 490-493.
\bibitem{Malykh}
Malykh S., Bakhanova Y., Kazakov A., Pusuluri K., Shilnikov A. Homoclinic chaos in the Rössler model //Chaos: An Interdisciplinary Journal of Nonlinear Science. – 2020. – v. 30(11). – p. 113126.
\bibitem{Obodan}
Obodan N. I., Adlutskii V. Y., Gromov V. A. Vulnerability assessment of loaded thin-walled shells under an external pulse action //Strength of Materials. – 2017. – v. 49(2). – p. 335-342.
\bibitem{Pipin}
Pipin V. V., Ragulskaya M. V., Chibisov S. M. Models of reactions of human heart as nonlinear dynamic system to cosmic and geophysical factors //Bulletin of experimental biology and medicine. – 2010. – v. 149(4). – p. 490-494.
\bibitem{Roessler}
Rössler O. E. An equation for continuous chaos //Physics Letters A. – 1976. – v. 57(5) – p. 397-398.
\bibitem{Safonov}
Safonov D. A., Klinshov V. V., Vanag V. K. Dynamical regimes of four oscillators with excitatory pulse coupling //Physical Chemistry Chemical Physics. – 2017. – v. 19(19). – p. 12490-12501.
\bibitem{Shilnikov}
Shilnikov A. Complete dynamical analysis of a neuron model //Nonlinear Dynamics. – 2012. – v. 68(3) – p. 305-328.
\bibitem{Stanoev}
Stanoev A., Nandan A. P., Koseska A. Organization at criticality enables processing of time‐varying signals by receptor networks //Molecular systems biology. – 2020. – v. 16(2) – p. e8870.

\bibitem{Tij}
Tij M., Santos A. Perturbation analysis of a stationary nonequilibrium flow generated by an external force //Journal of statistical physics. – 1994. – v. 76(5). – p. 1399-1414.
\bibitem{Ullmann}
Ullmann K., Caldas I. L. Transitions in the parameter space of a periodically forced dissipative system //Chaos, Solitons \& Fractals. – 1996. – v. 7(11). – p. 1913-1921.

\bibitem{Zi-zhen}
Zi-zhen L., Wan-xiong W., Cai-lin X. Dynamic model of crop growth system and numerical simulation of crop growth process under the multi-environment external force action //Applied Mathematics and Mechanics. – 2003. – v. 24(6). – p. 727-737.








\end{thebibliography}
\end{document}